\documentclass[preprint,aps,prd,amsmath,amssymb]{revtex4}
\usepackage{graphicx}
\usepackage{color}
\newcommand{\beq}{\begin{eqnarray}}
\newcommand{\eeq}{\end{eqnarray}}

\newcommand{\bmp}{\noindent\begin{minipage}{16cm}}
\newcommand{\emp}{\end{minipage}\vskip 7mm} 

\usepackage{graphicx}
\usepackage{dcolumn}
\usepackage{bm}
\usepackage{amsmath}
\usepackage{amsfonts}
\usepackage{bbm}
\usepackage{subfigure}
\usepackage{pxfonts}

\usepackage{ulem}

\newcommand{\drawsquare}[2]{\hbox{%
\rule{#2pt}{#1pt}\hskip-#2pt
\rule{#1pt}{#2pt}\hskip-#1pt
\rule[#1pt]{#1pt}{#2pt}}\rule[#1pt]{#2pt}{#2pt}\hskip-#2pt
\rule{#2pt}{#1pt}}
\newcommand{\Yfund}{\raisebox{-.5pt}{\drawsquare{6.5}{0.4}}}
\newcommand{\Ysymm}{\Yfund\hskip-0.4pt%
                    \Yfund}
\def\symm{\Ysymm}

\def\drawbox#1#2{\hrule height#2pt
        \hbox{\vrule width#2pt height#1pt \kern#1pt
              \vrule width#2pt}
              \hrule height#2pt}
\def\Fund#1#2{\vcenter{\vbox{\drawbox{#1}{#2}}}}
\def\Asym#1#2{\vcenter{\vbox{\drawbox{#1}{#2}
              \kern-#2pt 
              \drawbox{#1}{#2}}}}

\def\fund{\Fund{6.4}{0.3}}
\def\asymm{\Asym{6.4}{0.3}}

\begin{document}
\title{\Large  \color{red}  Conformal House
}
\author{Thomas A. {\sc Ryttov}$^{\color{blue}{\clubsuit}}$}
\email{ryttov@nbi.dk}
\author{Francesco {\sc Sannino}$^{\color{blue}{\varheartsuit}}$}
\email{sannino@ifk.sdu.dk} \affiliation{
$^{\color{blue}{\clubsuit}}$Niels Bohr Institute, Blegdamsvej 17, DK-2100 Copenhagen, Denmark \\
$^{\color{blue}{\varheartsuit}}${ CP}$^{ \bf 3}${-Origins}, 
Campusvej 55, DK-5230 Odense M, Denmark.\footnote{{ C}entre of Excellence for { P}article { P}hysics { P}henomenology devoted to the understanding of the {Origins} of Mass in the universe. This is the new affiliation from September 1$^{st}$ 2009.}}
\begin{flushright}
{\it CP$^3$- Orgins: 2009-2}
\end{flushright}
\begin{abstract}
We investigate the gauge dynamics of nonsupersymmetric SU(N) gauge theories featuring the simultaneous presence of fermionic matter transforming according to two distinct representations of the underlying gauge group. We bound the regions of flavors and colors which can yield a physical infrared fixed point. As a consistency check we recover the previously investigated conformal windows bounds when restricting to a single matter representation. The earlier conformal windows can be imagined to be part now of the new conformal {\it house}. We predict the nonperturbative anomalous dimensions at the infrared fixed points.  We further investigate the effects of adding mass terms to the condensates on the conformal house chiral dynamics and construct the simplest instanton induced effective Lagrangian terms. \end{abstract}

\maketitle
\newpage
\tableofcontents
\newpage

\section{Introduction}

Models of dynamical electroweak symmetry breaking constitute one of the best motivated extensions of the Standard Model (SM).  A sensible model building requires, however, a deep knowledge of gauge dynamics in a regime where perturbation theory fails.  It is, hence, of utmost importance to gain information on the nonperturbative dynamics of non-abelian four dimensional gauge theories.

Recent studies of the dynamics of gauge theories featuring fermions transforming according to higher dimensional representations of the new gauge group led to several interesting phenomenological possibilities \cite{Sannino:2004qp,Dietrich:2005jn,Dietrich:2006cm} such as Minimal Walking Technicolor (MWT) \cite{Foadi:2007ue} and Ultra Minimal Walking Technicolor (UMT) \cite{Ryttov:2008xe}. We observe that higher dimensional representations have been used earlier in particle physics phenomenology and time honored examples are grand unified models. The initial discovery \cite{Sannino:2004qp} that theories with fermions transforming according to higher dimensional representations develop an infrared fixed point (IRFP) for an extremely small number of flavors and colors has been tested using several analytic methods  \cite{Sannino:2004qp,Dietrich:2006cm,Ryttov:2007cx} for $SU(N)$ gauge groups.  The analysis for $SO(N)$ ad $Sp(2N)$ gauge groups has been performed in \cite{Sannino:2009aw}. This discovery is important since it permits the construction of several explicit UV-complete models able to break the electroweak symmetry dynamically while naturally featuring small contributions to the electroweak precision parameters \cite{Appelquist:1998xf,Kurachi:2006mu,Foadi:2007ue}. Simultaneously it also helps alleviating the Flavor Changing Neutral Currents while the models also feature explicit candidates of asymmetric dark matter \cite{Foadi:2007ue,Ryttov:2008xe}. These models are economical since they require the introduction of a very small number of underlying elementary fields and can feature a light composite Higgs \cite{Dietrich:2005jn,Dietrich:2006cm,Hong:2004td}.  Recent analyses lend further support to the latter observation \cite{Doff:2009nk,Doff:2008xx,Doff:2009kq}.

If a theory develops an IRFP fixed point then at large distances displays a conformal behavior. One can envision several ways to depart from conformality. {} For example one can add a relevant operator such as an explicit fermion mass term  or decrease the number of flavors. If the departure from conformality is {\it soft}, meaning that the IRFP is quasi-reached the gauge coupling constant runs slowly over a long range of energies and the theory is said to   {\it walk}
\cite{Eichten:1979ah,Holdom:1981rm,Yamawaki:1985zg,Appelquist:1986an}. This is, however, not the best way to define a walking theory since the coupling constant is not a physical quantity. In fact one should look at two and higher point correlators and determine the associated scaling exponents. In a (quasi)-conformal theory the correlators will have a characteristic power law behavior.
Gauge theories developing an IRFP are natural ultraviolet completions of unparticle \cite{Georgi:2007ek,Georgi:2007si,Georgi:2008pq,Georgi:2009xq} models \cite{Sannino:2008nv,Sannino:2008ha}. The effects of the instantons and their interplay with the fermion-mass operator on the conformal window have been evaluated in \cite{Sannino:2008pz}.  Within the SD approach these effects were investigated in \cite{Appelquist:1997dc}.

Non-abelian gauge theories exist in a number of  distinct phases which can be classified according to the characteristic dependence of the potential energy on the distance between
two well separated static sources.  The collection of all of these different behaviors,
when represented, for example, in the  flavor-color space, constitutes the {\it phase diagram} of the given gauge theory. The phase diagram of $SU(N)$ gauge theories as functions of number of flavors, colors and matter representation has been investigated in \cite{Sannino:2004qp,Dietrich:2006cm,Ryttov:2007sr,Ryttov:2007cx,Sannino:2008ha}. Interesting applications have been envisioned not only for the LHC phenomenology \cite{Sannino:2004qp,Foadi:2007ue,Belyaev:2008yj,Christensen:2005cb,Gudnason:2006mk,Dietrich:2009ix} but also for Cosmology \cite{Nussinov:1985xr,Barr:1990ca,Foadi:2008qv,Ryttov:2008xe,Nardi:2008ix, Gudnason:2006yj,Kainulainen:2006wq,Kouvaris:2007iq,Cline:2008hr,Kouvaris:2008hc,Kikukawa:2007zk,Jarvinen:2009wr,Antipin:2009ch}. The nonperturbative dynamics of these models is being investigated via first principles lattice computations by several groups  \cite{Catterall:2007yx,Catterall:2008qk,
Shamir:2008pb,DelDebbio:2008wb,DelDebbio:2008zf, Hietanen:2008vc,Hietanen:2008mr,Appelquist:2007hu,Deuzeman:2008sc,Fodor:2008hn,DelDebbio:2008tv,DeGrand:2008kx,Appelquist:2009ty,Hietanen:2009az,Deuzeman:2009mh,Lucini:2007sa}. In the literature the reader can also find various attempts to gain information on the nonperturbative gauge dynamics  using  gauge-gravity type duality and we cite here only a few recent efforts \cite{Hirn:2008tc,Dietrich:2008ni,Nunez:2008wi,Mintakevich:2009wz}.  We have also extended the analysis of  the zero temperature and matter density phase diagram to $SO(N)$ and $Sp(2N)$ gauge theories with matter in a single matter representation \cite{Sannino:2009aw}. An interesting universal picture emerges unifying the phase diagrams of the various gauge groups.

Till now the various investigations dealt with fermions in a single representation of the gauge group. In fact these constitute only a small fraction of all of the possible gauge theories we can envision built out of fermions in several representations. A priori there is no reason to exclude these theories from interesting applications. In fact, we have very recently shown that one of these theories leads to a novel model of dynamical electroweak symmetry breaking possessing several interesting phenomenological features \cite{Ryttov:2008xe}.

The goal of this paper is to initiate the first systematic study of conformal gauge dynamics associated to nonsupersymmetric gauge theories featuring matter in two different representations of the undelying gauge group. The region in flavor/color space bounding the fraction of the theory developing a conformal behavior at large distances is a three-dimensional volume. Two faces of this volume correspond the the conformal areas of the gauge theory when on of the flavor numbers is set to zero. These areas are often referred as conformal {\it windows}. It is then natural to indicate the conformal volumes as the conformal {\it houses} whose {\it windows} are the two dimensional conformal areas.

What methods can we use to draw the boundary of the conformal houses? 
The ideal, would be, to use  first principle lattice simulations. Although this is possible, in principle, it is simply too expensive to unveil the entire phase diagram and one can only explore a small portion of it. This is especially true for gauge theories with multiple representations given that the phase diagram is multidimensional. Hence, before launching into a lattice simulation, one should have an idea of how the phase diagram looks like in order to select the relevant theories to investigate. We have considered different analytic methods in our previous studies of the conformal windows. The following table, taken from \cite{Sannino:2009aw}, neatly summarizes the range of applicability of the three most used methods. They are  the all-orders beta function (BF) \cite{Ryttov:2007cx}; The truncated Schwinger-Dyson equation (SD) \cite{Appelquist:1988yc,Cohen:1988sq,Miransky:1996pd} (referred also as the ladder approximation in the literature);  The Appelquist-Cohen-Schmaltz (ACS) conjecture \cite{Appelquist:1999hr} which makes use of the counting of the thermal degrees of freedom at high and low temperature. In the Table below we compared directly the various analytical methods. 
\begin{table}[ht]
\caption{Direct comparison among the various analytic methods}
\centering
\begin{tabular}
{|c|c|c|c|c|c|} \hline
{Method} &  ~~~Fund. - Rep. ~~~& ~Higher Rep.~&~Multiple Rep.~&~~Susy~~& ~~~$\gamma$~~~  \\ \hline \hline
BF & +  & +  &+ & + & +  \\
SD &  + & +  & - & - & -\\
ACS & + & - & - & + &  - \\
 \hline 
 \end{tabular} 
 \label{comparison}
\end{table}
 The three plus signs in the second column indicate that the three analytic methods do constrain the conformal window of $SU$, $Sp$ and $SO$ gauge theories with fermions in the fundamental representation. Only BF and SD provide useful constraints in the case of the higher dimensional representations as summarized in the third column. {}When multiple representations participate in the gauge dynamics the BF constraints can be used directly \cite{Ryttov:2007cx,Ryttov:2008xe} to determine the extension of the conformal (hyper)volumes while extra dynamical information and approximations are required in the $SD$ approach. Since gauge theories with fermions in several representations of the underlying gauge group must contain higher dimensional representations the ACS is less efficient in this case \footnote{We do not consider super QCD a theory with higher dimensional representations.}. These results are summarized in the fourth column. The all-orders beta function reproduces the supersymmetric exact results when going over the super Yang-Mills case, the ACS conjecture was  proved successful when tested against the supersymmetric conformal window results \cite{Appelquist:1999hr}. However the SD approximation does not reproduce any supersymmetric result \cite{Appelquist:1997gq}.  The results are summarized in the fifth column. Finally, it is of  theoretical and phenomenological interest -- for example to construct sensible UV completions of models of dynamical electroweak symmetry breaking and unparticles --  to compute the anomalous dimension of the mass of the fermions at the (near) conformal fixed point.  Only the all-orders beta function provides a simple closed form expression as it is summarized in the sixth column.

The table above clearly shows, by direct comparison of the various analytic method used to analyze the phase diagrams in \cite{Sannino:2009aw}, that the conjectured all-orders bera function for nonsupersymmetric gauge theories with fermionic matter in arbtriray representations of the gauge group  \cite{Ryttov:2007cx} is the most efficient method. We will henceforth use the BF method.

We will then investigate the effects of adding relevant mass operators on the chiral dynamics. {}Following \cite{Sannino:2008pz} we will also estimate the effects of the instanton induced interactions on the chiral dynamics. 

Our results greatly enlarge the number of gauge theories which can be used to extend the SM of particle interactions. 

\section{Turning the Window Upside Down}

The conformal window of non-abelian gauge theories containing matter transforming according to a single specific representation of the gauge group has received much attention throughout many years \cite{Appelquist:1988yc,Cohen:1988sq,Miransky:1996pd,Appelquist:1999hr,Sannino:2004qp,Dietrich:2006cm,Ryttov:2007cx,Sannino:2009aw}. In the past many different theoretical tools have been developed to estimate its size and shape.
By direct comparison of the predictions stemming from the various methods it has been shown in  \cite{Sannino:2009aw}  that the recently conjectured beta function \cite{Ryttov:2007cx} for nonsupersymmetric gauge theories with fermionic matter allows for useful bounds of the various conformal windows and yields the maximum amount of information. The predictions are unambiguous and can be tested. The conjecture is non-trivially supported by all the recent lattice data \cite{Catterall:2007yx,DelDebbio:2008wb,Catterall:2008qk,Appelquist:2007hu,
Shamir:2008pb,Deuzeman:2008sc,Lucini:2007sa}.

Considering an $SU(N)$ gauge group (the generalization to $SO$ and $Sp$ groups appeared in \cite{Sannino:2009aw}) with $N_f(r_i)$ Dirac flavors belonging to the representation $r_i,\ i=1,\ldots,k$ of the gauge group the proposed beta function reads
\begin{eqnarray}
\beta(g) &=&- \frac{g^3}{(4\pi)^2} \frac{\beta_0 - \frac{2}{3}\, \sum_{i=1}^k T(r_i)\,N_{f}(r_i) \,\gamma_i(g^2)}{1- \frac{g^2}{8\pi^2} C_2(G)\left( 1+ \frac{2\beta_0'}{\beta_0} \right)} \ ,
\end{eqnarray}
with
\begin{eqnarray}
\beta_0 =\frac{11}{3}C_2(G)- \frac{4}{3}\sum_{i=1}^k \,T(r_i)N_f(r_i) \qquad \text{and} \qquad \beta_0' = C_2(G) - \sum_{i=1}^k T(r_i)N_f(r_i)  \ .
\end{eqnarray}

The generators $T_r^a,\, a=1\ldots N^2-1$ of the gauge group in the
representation $r$ are normalized according to
$\text{Tr}\left[T_r^aT_r^b \right] = T(r) \delta^{ab}$ while the
quadratic Casimir $C_2(r)$ is given by $T_r^aT_r^a = C_2(r)I$. The
trace normalization factor $T(r)$ and the quadratic Casimir are
connected via $C_2(r) d(r) = T(r) d(G)$ where $d(r)$ is the
dimension of the representation $r$. The adjoint
representation is denoted by $G$. For the reader's convenience we list in Table \ref{factors} the
explicit group factors for the representations used here. A complete
list of all of the group factors for any representation and the way
to compute them is available in Table II of \cite{Dietrich:2006cm}
and the associated appendix \footnote{The normalization for the
generators here is different than the one adopted in
\cite{Dietrich:2006cm}.}.  {}For $Sp$ and $SO$ gauge groups one can read off the group factors in \cite{Sannino:2009aw}.

\begin{table}
\begin{center}
    \begin{tabular}{c||ccc }
    r & $ \quad T(r) $ & $\quad C_2(r) $ & $\quad
d(r) $  \\
    \hline \hline
    $ \fund $ & $\quad \frac{1}{2}$ & $\quad\frac{N^2-1}{2N}$ &\quad
     $N$  \\
        $\text{$G$}$ &\quad $N$ &\quad $N$ &\quad
$N^2-1$  \\
        $\symm$ & $\quad\frac{N+2}{2}$ &
$\quad\frac{(N-1)(N+2)}{N}$
    &\quad$\frac{N(N+1)}{2}$    \\
        $\asymm$ & $\quad\frac{N-2}{2}$ &
    $\quad\frac{(N+1)(N-2)}{N}$ & $\quad\frac{N(N-1)}{2}$
    \end{tabular}
    \end{center}
\caption{Relevant group factors for the representations used
throughout this paper. However, a complete list of all the group
factors for any representation and the way to compute them is
available in Table II and the appendix of
\cite{Dietrich:2006cm}.}\label{factors}
    \end{table}

One should note that the beta function is given in terms of the anomalous dimension of the fermion mass $\gamma=-\frac{d\ln m}{d\ln \mu}$ where $m$ is the renormalized mass, similar to the supersymmetric case \cite{Novikov:1983uc,Shifman:1986zi,Jones:1983ip}. Indeed the construction of the above beta function is inspired by the one of their supersymmetric cousin theories. At small coupling it coincides with the two-loop beta function and in the non-perturbative regime reproduces earlier known exact results. Similar to the supersymmetric case it allows for a bound of the conformal window \cite{Seiberg:1994pq}. In the supersymmetric case where additional checks can be made the bound is actually believed to give the true conformal window. We stress that the predictions of the conformal window coming from the above beta function are nontrivially supported by all the recent lattice results \cite{Catterall:2007yx,DelDebbio:2008wb,Catterall:2008qk,Appelquist:2007hu,
Shamir:2008pb,Deuzeman:2008sc,Lucini:2007sa}.
\begin{figure}[h!]
\includegraphics[height=7cm,width=10cm]{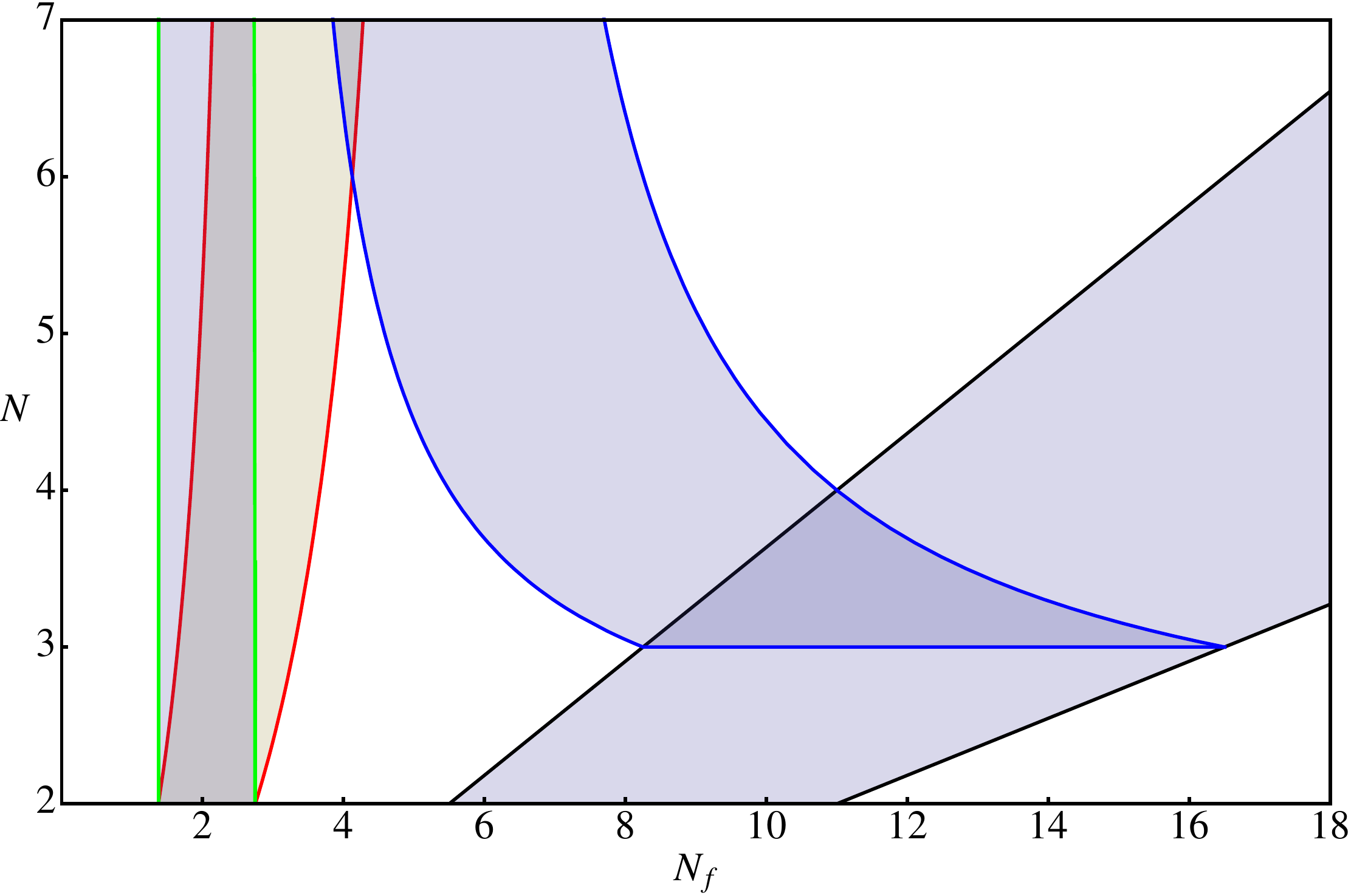}
\caption{Phase diagram for non-supersymmetric theories with fermions in the: i) fundamental representation (black), ii) two-indexed antisymmetric representation (blue), iii) two-indexed symmetric representation (red), iv) adjoint representation (green) as a function of the number of colors and number of flavors. The shaded areas depict the corresponding conformal windows stemming from the all-orders beta function. To the right of the shaded ares the theories are no longer asymptotically free while to the left of the shaded areas the theories are expected to break chiral symmetry. }\label{ReversePD}
\end{figure}

Let us review the case of a single representation. The loss of asymptotic freedom occurs when the first coefficient of the beta function changes sign
\begin{eqnarray}\label{SingleAF}
11C_2(G) &=& 4 T(r) N_f(r) \ .
\end{eqnarray}
For a given representation this determines the critical number of colors as a function of the number of flavors at which asymptotic freedom is lost. Second we note that at the zero of the beta function we have
\begin{eqnarray}
\gamma &=& \frac{11C_2(G) - 4T(r)N_f(r)}{2T(r)N_f(r)} \ .
\end{eqnarray}
Therefore specifying the value of the anomalous dimensions at the infrared fixed point yields the last constraint needed to construct the conformal window. Having reached the zero of the beta function the theory is conformal in the infrared. For a theory to be conformal the dimension of the non-trivial spinless operators must be larger than one in order to not contain negative norm states \cite{Mack:1975je,Flato:1983te,Dobrev:1985qv}.  Since the dimension of the chiral condensate is $3-\gamma$ we see that $\gamma = 2$ yields the maximum possible bound \footnote{Note that $\gamma \leq 2$ is an {\it exact} bound  \cite{Mack:1975je,Flato:1983te,Dobrev:1985qv}, i.e. does not depend on model computations. If it turns out that dynamically a smaller value of $\gamma$ actually delimits the conformal window this value must be less than $2$ and hence does not affect our results on the bound of the conformal windows.  }
\begin{eqnarray}\label{SingleBound}
11C_2(G) &=& 8T(r) N_f(r) \ .
\end{eqnarray}
For a given representation this determines the critical number of color as a function of the minimum number of flavors for which an infrared fixed point can be reached.
At this point one would have reported the graphical representation of the various conformal windows by keeping the number of flavors on the vertical axis and the number of colors on the horizontal axis. Given that we have in mind to generalize this bi-dimensional representation to the case of one more representation it is more convenient to draw it from the beginning with the number of colors on the vertical axis. This is summarized
 in Fig. \ref{ReversePD} where we used equations \eqref{SingleAF} and \eqref{SingleBound}.

\section{ Moving in the House}\label{moving-in}

Let us now generalize to multiple representations. First, the loss of asymptotic freedom is determined by the change of sign in the first coefficient of the beta function. This occurs when
\begin{eqnarray} \label{MultiAF}
\sum_{i=1}^{k} \frac{4}{11} T(r_i) N_f(r_i) = C_2(G) \ .
\end{eqnarray}
Second, we note that at the zero of the beta function we have
\begin{eqnarray}
\sum_{i=1}^{k} \frac{2}{11}T(r_i)N_f(r_i)\left( 2+ \gamma_i \right) = C_2(G) \ .
\end{eqnarray}

 Therefore specifying the value of the anomalous dimensions at the infrared fixed point yields the last constraint needed to construct the conformal region. Having reached the zero of the beta function the theory is conformal in the infrared. For a theory to be conformal the dimension of the non-trivial spinless operators must be larger than one in order to not contain negative norm states \cite{Mack:1975je,Flato:1983te,Dobrev:1985qv}.  Since the dimension of each chiral condensate is $3-\gamma_i$ we see that $\gamma_i = 2$, for all representations $r_i$, yields the maximum possible bound
\begin{eqnarray}\label{MultipleBound}
\sum_{i=1}^{k} \frac{8}{11} T(r_i)N_f(r_i) = C_2(G) \ .
\end{eqnarray}

\begin{figure}[ht!]
{\includegraphics[height=6.5cm,width=8.82cm]{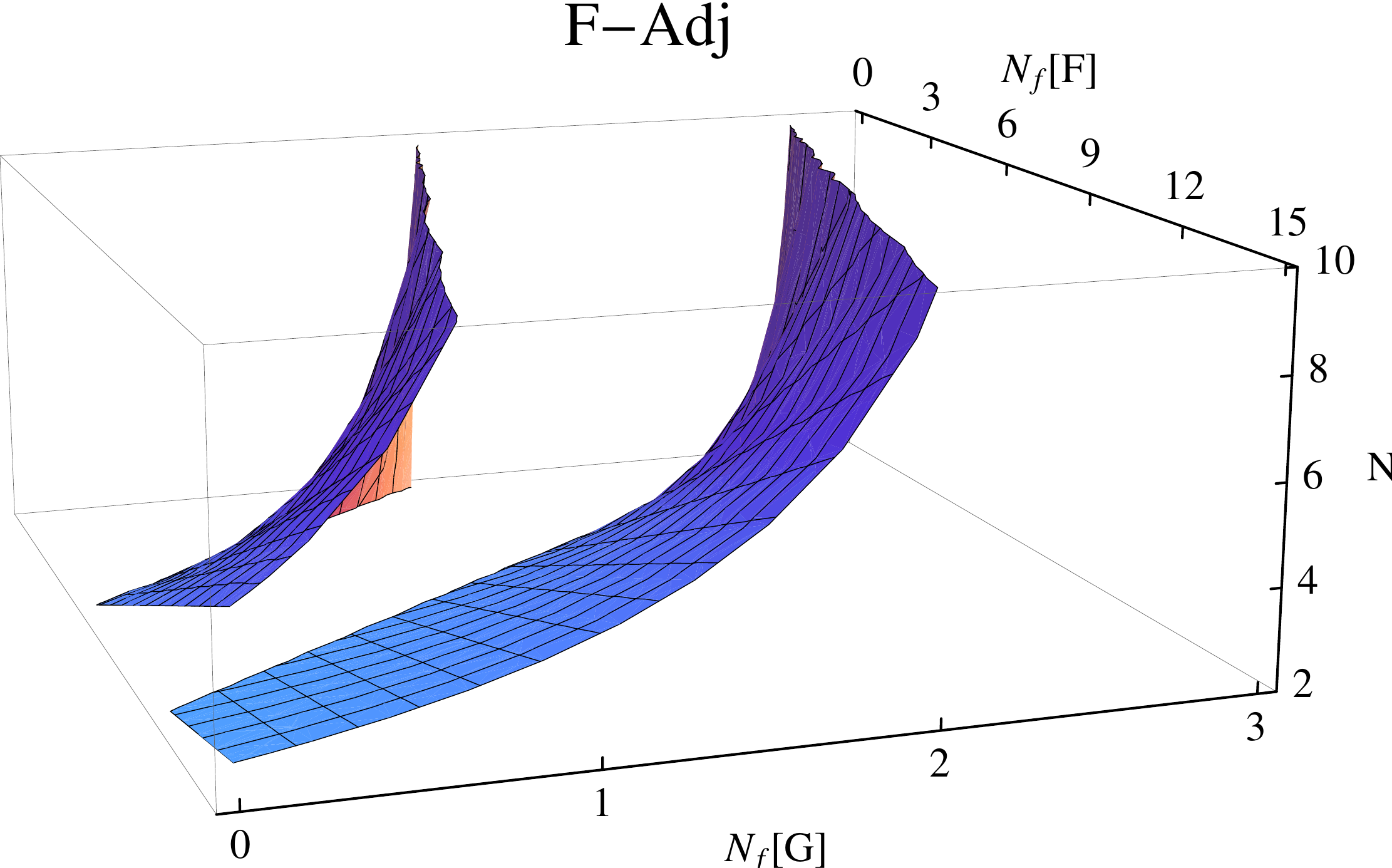}\hspace{0.8cm}}
\caption{The conformal {\it house} for a non-supersymmetric gauge theory containing fundamental and adjoint fermions. To the right of the right surface the theories are non-asymptotically free while to the left of the left surface the theories break chiral symmetry. Between the two surfaces the theories can develop an infrared fixed point.}
\label{F-Adj}
\end{figure}

\begin{figure}[h]
{\includegraphics[height=6.5cm,width=7.70cm]{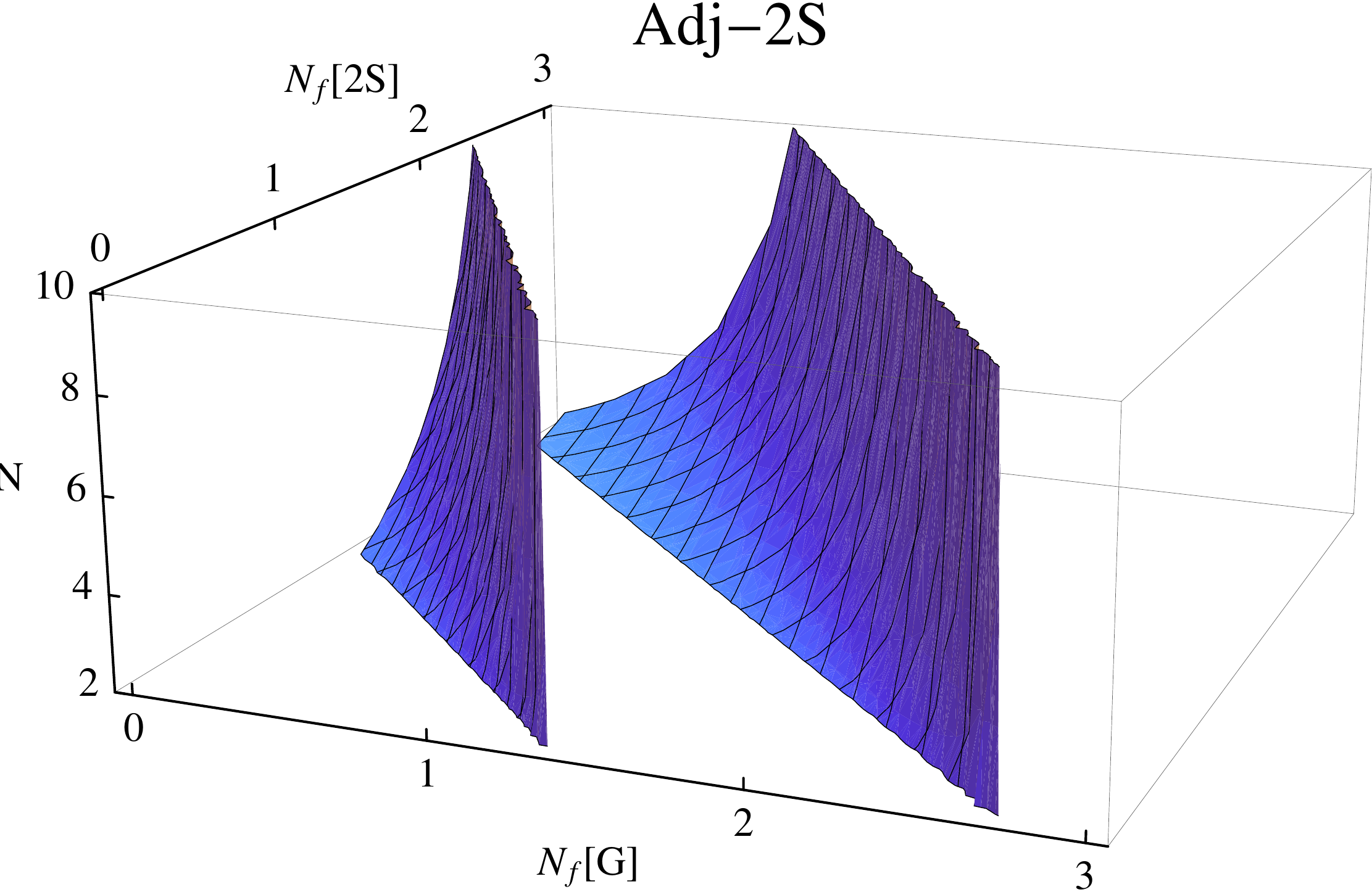}\hspace{0.8cm}\includegraphics[height=6.5cm,width=7.70cm]{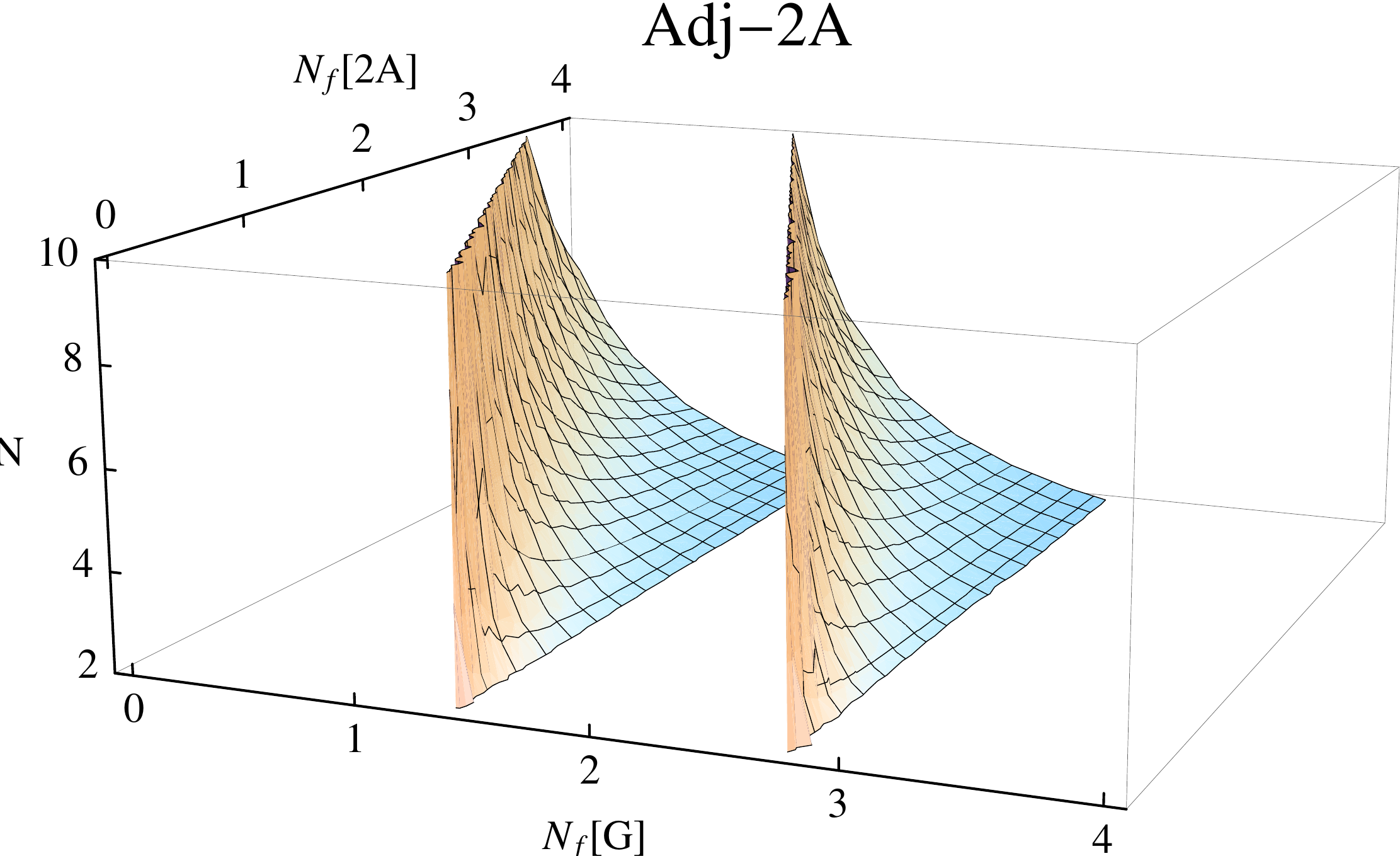}}
\caption{The conformal {\it house} for a non-supersymmetric gauge theory containing fermions in the adjoint and two-indexed symmetric representations (left) and in the adjoint and two-indexed antisymmetric representations (right).}\label{fig:adj}
\end{figure}

For two distinct representations the conformal region is a three dimensional volume, i.e. the conformal {\it house}. The {\it windows} of the house correspond
exactly to the conformal windows presented in the previous section.  In Fig. \ref{F-Adj} we plot the bound of the conformal volume in the case of fundamental and adjoint fermions.
{}For completeness we also plot below in Fig.~\ref{fig:fund} the bound on the conformal house with one species of fermions in the fundamental representation and the other in the two-index (anti)symmetric  in the (right)left panel.  We consider only two-index representations in Fig.~\ref{fig:adj}, more specifically we consider the adjoint representation together with the two-index (anti)symmetric in the \ref{fig:adj} (right) left panel. Note that to the right of the right surface the theories are non-asymptotically free while to the left of the left surface the theories break chiral symmetry. Between the two surfaces the theories can develop an infrared fixed point.
\begin{figure}[h]
{\includegraphics[height=6.5cm,width=7.70cm]{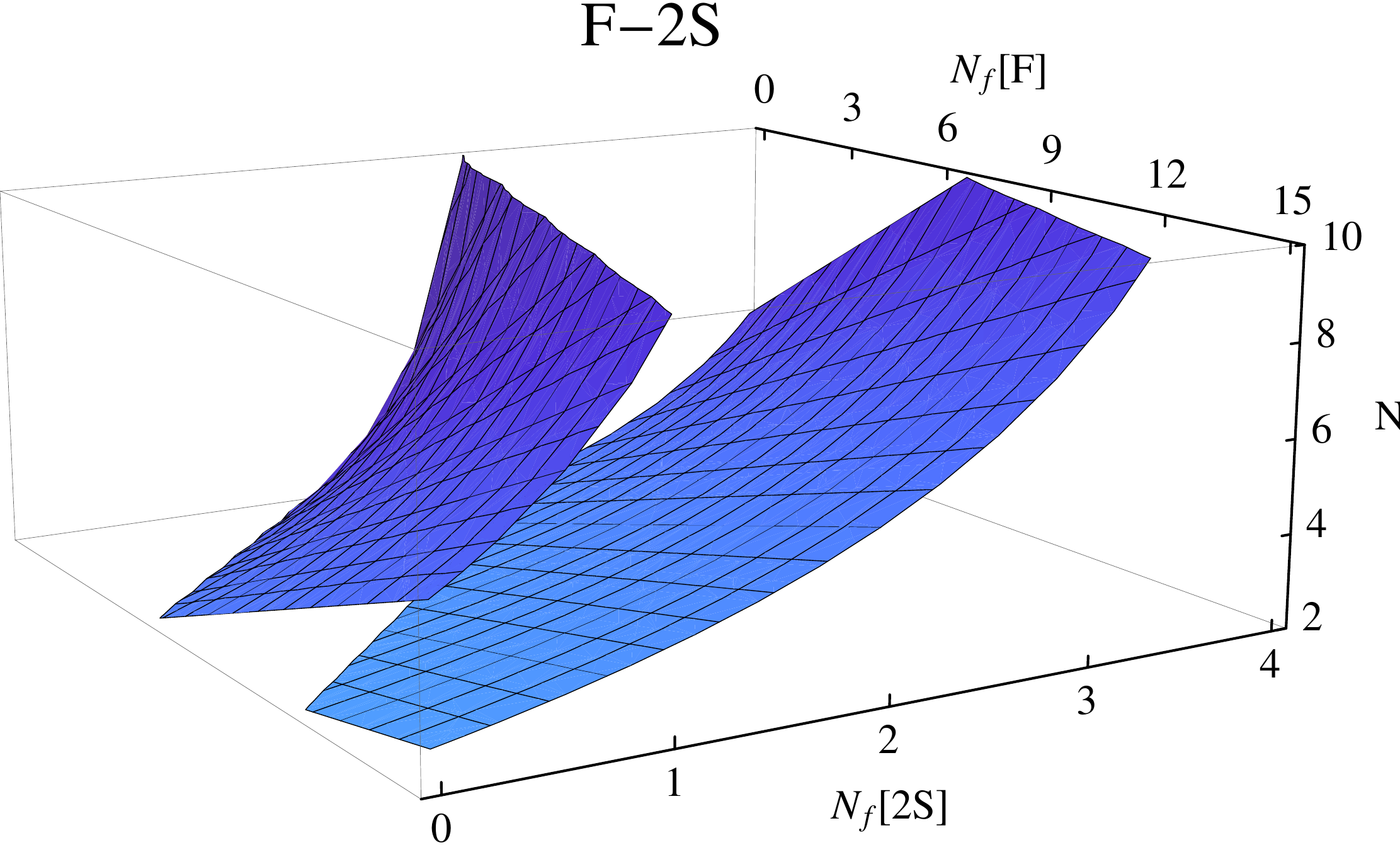}\hspace{0.8cm}\includegraphics[height=6.5cm,width=7.70cm]{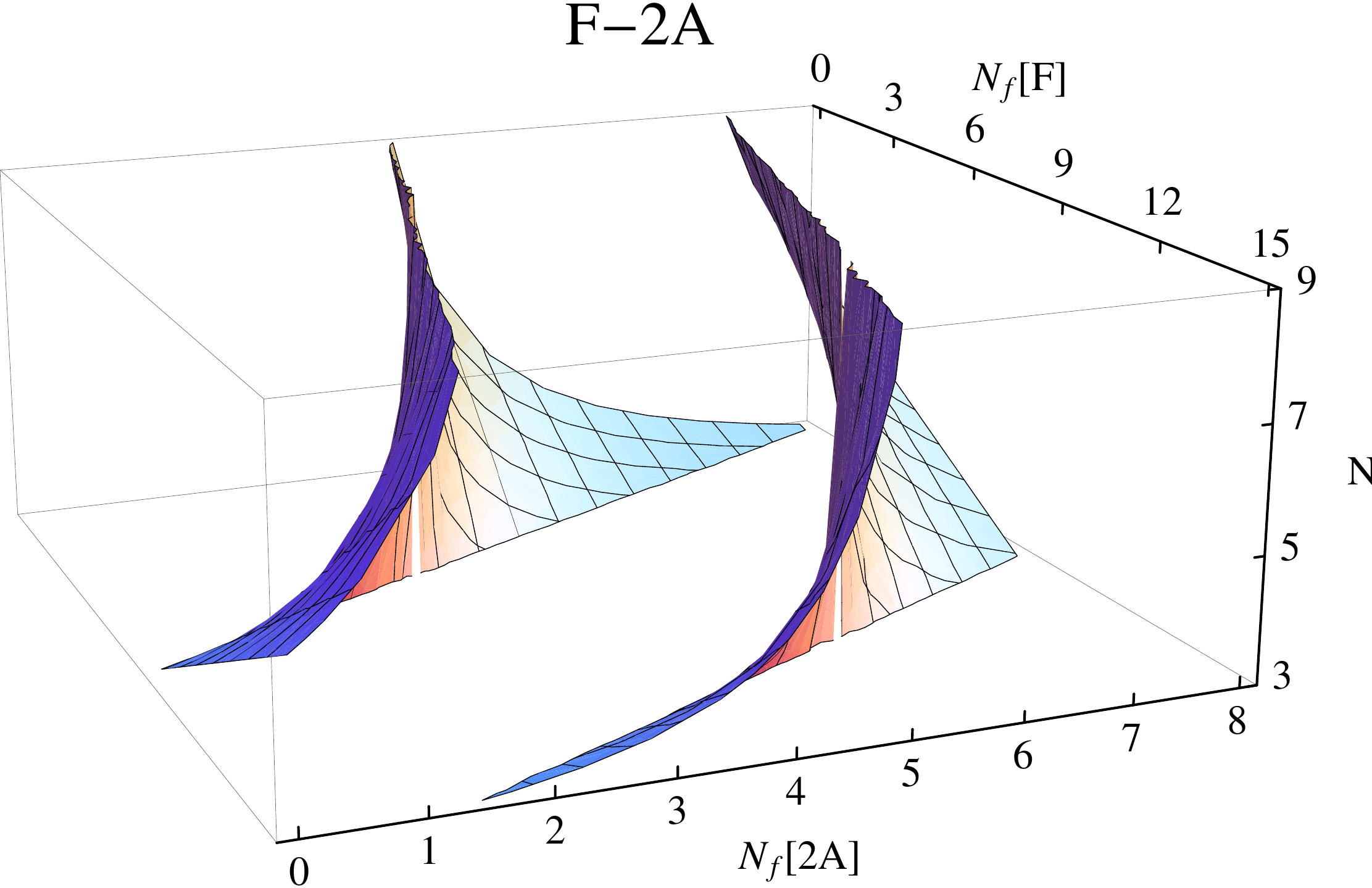}}
\caption{The conformal {\it house} for a non-supersymmetric gauge theory containing fermions in the fundamental and two-indexed symmetric representations (left) and in the fundamental and two-indexed antisymmetric representations (right).}\label{fig:fund}
\end{figure}
\begin{figure}[h]
{\includegraphics[height=6.5cm,width=7.70cm]{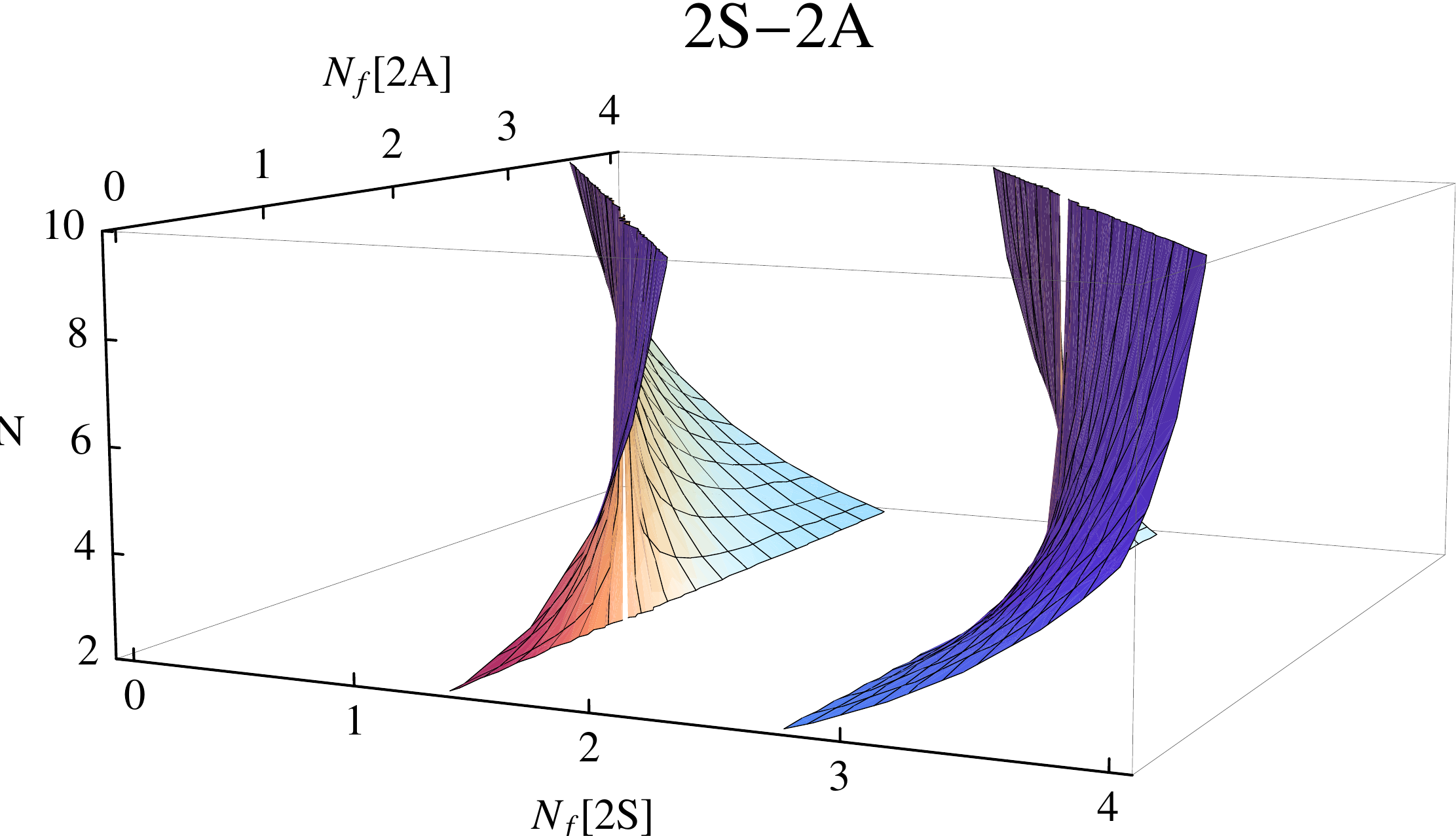}\hspace{0.8cm}}
\caption{The conformal {\it house} for a non-supersymmetric gauge theory containing fermions in the two-indexed symmetric and two-indexed antisymmetric representations.}\label{fig:two}
\end{figure}
Finally in Fig.~\ref{fig:two} we consider the last case in which one representation corresponds to the two-index symmetric and the other one is the two-index antisymmetric.

\section{Moving out}

Consider again an $SU(N)$ gauge theory with $N_f(r_{\psi})$ and $N_f(r_\xi)$ fermionic Dirac flavors in the representation $r_{\psi}$ and $r_\xi$ respectively. Also assume that we have chosen the number of flavors in such a way that the theory is conformal in the infrared. Let us first define with $\Lambda_U$ the scale for which the coupling constant, as we increase the energy, drops by $2/3$ of the fixed point value.

Following the single representation case \cite{Sannino:2008pz} we shall add a small mass term for each of the fermions and study the behavior of the associated condensates. Denote by $\psi_{c_{\psi}}^{f_{\psi}}$, $\tilde{\psi}^{c_{\psi}}_{f_{\psi}}$, $\xi_{c_{\xi}}^{f_{\xi}}$ and $\tilde{\xi}^{c_{\xi}}_{f_{\xi}}$ four left transforming Weyl fermions. They are the components of the two Dirac fermions $(\psi,\bar{\tilde{\psi}})$ and $(\xi,\bar{\tilde{\xi}})$ belonging to the representations $r_{\psi}$ and $r_{\xi}$ respectively. Here $c_{\psi}$ and $c_{\xi}$ represent the color indices while $f_{\psi}$ and $f_{\xi}$ represent the flavor indices. The two mass operators are:
\begin{eqnarray}
-m_{\psi} \Lambda_U^{\gamma_{\psi}} \text{Tr}\left[ \mathcal{O}_{\tilde{\psi}\psi} \right]
-m_{\xi} \Lambda_U^{\gamma_{\xi}} \text{Tr}\left[ \mathcal{O}_{\tilde{\xi}\xi} \right] + \text{h.c.} \ ,
\end{eqnarray}
with $m_{\psi}$ and $m_{\xi}$ being the respective fermion masses and
\begin{eqnarray}
\mathcal{O}_{\tilde{\psi}\psi}{}_{f_{\psi}}^{f_{\psi}'} = \tilde{\psi}^{f_{\psi}'}\psi_{f_{\psi}} \ , \qquad \mathcal{O}_{\tilde{\xi}\xi}{}_{f_{\xi}}^{f_{\xi}'} = \tilde{\xi}^{f_{\xi}'}\xi_{f_{\xi}} \ ,
\end{eqnarray}
are gauge singlet operators of dimension $d_{\tilde{\psi}\psi} = 3- \gamma_{\psi}$ and $d_{\tilde{\xi}\xi} = 3- \gamma_{\xi}$ respectively. Also $\gamma_{\psi}$ and $\gamma_{\xi}$ are the anomalous dimensions of the respective mass terms. As shown above they are both positive numbers less than two at the fixed point. Note that it is reasonable to keep the leading term in $m/\Lambda_{U} $ since we are working in the small mass approximation.

We first define the following two flavor singlet operators $\mathcal{O}_{\psi}$ and $\mathcal{O}_{\xi}$ via $\text{Tr}\left[ \mathcal{O}_{\tilde{\psi}\psi} \right] = N_f(r_{\psi}) \mathcal{O}_{\psi}$ and $\text{Tr}\left[ \mathcal{O}_{\tilde{\xi}\xi} \right] = N_f(r_{\xi}) \mathcal{O}_{\xi}$. For a conformal theory there are no mass terms and the spectrum is continuous. However having explicitly introduced small quark masses conformal and chiral symmetry break and a discrete spectrum emerges. We expand $\langle O_{\psi} \rangle$ and $\langle O_{\xi} \rangle$ in terms of orthornormal single particle states
\begin{eqnarray}
\mathcal{O}_{\psi}(x) = \sum_{n=1}^{\infty} a_n \delta_n(x) \ , \qquad \mathcal{O}_{\xi}(x) = \sum_{n=1}^{\infty} b_n \rho_n(x) \ .
\end{eqnarray}
Let the discrete tower of massive states be dictated by two mass gaps $\Delta_{\psi}$ and $\Delta_{\xi}$ in such a way that the spectrum is spaced according to
\begin{eqnarray}
M_{\psi,n}^2 = \Delta_{\psi}^2 n \ ,\qquad M_{\xi,n}^2 = \Delta_{\xi}^2 n \ ,
\end{eqnarray}
with $n$ an integer number \cite{Stephanov:2007ry}. One can choose another spectral decomposition without altering our results \cite{Stephanov:2007ry,Sannino:2008nv,Sannino:2009aw}.
The expansion coefficients can then be written as
\begin{eqnarray}
a_n^2 = \mathcal{F}_{d_{\tilde{\psi}\psi}} \Delta_{\psi}^2 \left( M_{\psi,n}^2 \right)^{d_{\tilde{\psi}\psi}-2} \ , \qquad b_n^2 = \mathcal{F}_{d_{\tilde{\xi}\xi}} \Delta_{\xi}^2 \left( M_{\xi,n}^2 \right)^{d_{\tilde{\xi}\xi}-2}
\end{eqnarray}
where $\mathcal{F}_{d_{\tilde{\psi}\psi}}$ and $\mathcal{F}_{d_{\tilde{\xi}\xi}}$ are functions depending on the dimensions of the two operators and the underlying gauge dynamics.
Having replaced the continuous spectrum with a tower of massive single particle states and due to the fictitious mass terms the potential becomes
\begin{eqnarray}\label{potential}
V &=& m_{\psi}\Lambda_U^{\gamma_{\psi}} N_f(r_{\psi}) \sum_{n=1}^{\infty}a_n\delta_n + m_{\xi}\Lambda_U^{\gamma_{\xi}} N_f(r_{\xi}) \sum_{n=1}^{\infty}b_n\rho_n + \text{h.c.} \nonumber \\
&& + \sum_{n=1}^{\infty} M_{\psi,n}^2 \bar{\delta}_n \delta_n + \sum_{n=1}^{\infty} M_{\xi,n}^2 \bar{\rho}_n \rho_n \ .
\end{eqnarray}
Minimizing the potential we find
\begin{eqnarray}
\langle \delta_n \rangle = -\bar{m}_{\psi} \Lambda_U^{\gamma_{\psi}} N_f(r_{\psi}) \frac{a_n}{M_{\psi,n}^2} \ , \qquad  \langle \rho_n \rangle = -\bar{m}_{\xi} \Lambda_U^{\gamma_{\xi}} N_f(r_{\xi}) \frac{b_n}{M_{\xi,n}^2} \ ,
\end{eqnarray}
from which we easily deduce
\begin{eqnarray}
\langle O_{\psi} \rangle &=& -\bar{m}_{\psi} \Lambda_U^{\gamma_{\psi}} N_f(r_{\psi}) \mathcal{F}_{d_{\tilde{\psi}\psi}} \sum_{n=1}^{\infty} \left(\Delta_{\psi}^2 n \right)^{-\gamma_{\psi}} \Delta_{\psi}^2 \ , \\
\langle O_{\xi} \rangle &=& -\bar{m}_{\xi} \Lambda_U^{\gamma_{\xi}} N_f(r_{\xi}) \mathcal{F}_{d_{\tilde{\xi}\xi}}  \sum_{n=1}^{\infty} \left(\Delta_{\xi}^2 n \right)^{-\gamma_{\xi}} \Delta_{\xi}^2 \ .
\end{eqnarray}

Note that we have written the two sums in a suggestive way. Taking the limit when the artificial mass gaps approach zero the sums become integrals which are evaluated to yield
\begin{eqnarray}
\langle \mathcal{O}_{\psi} \rangle &=& -\bar{m}_{\psi} \Lambda_U^{\gamma_{\psi}} N_f(r_{\psi}) \mathcal{F}_{d_{\tilde{\psi}\psi}} \Omega_{\psi}\left[\Lambda_{UV}, \Lambda_{IR} \right]\ , \\
\langle \mathcal{O}_{\xi} \rangle &=& -\bar{m}_{\xi} \Lambda_U^{\gamma_{\xi}} N_f(r_{\xi}) \mathcal{F}_{d_{\tilde{\xi}\xi}}  \Omega_{\xi}\left[\Lambda_{UV}, \Lambda_{IR} \right]\ ,
\end{eqnarray}
with
\begin{eqnarray}
\Omega_{i}\left[\Lambda_{UV}, \Lambda_{IR} \right] &=& \frac{1}{1-\gamma_{i}} \left[ \Lambda_{UV}^{2(1-\gamma_{i})}- \Lambda_{IR}^{2(1-\gamma_{i})} \right]  \ , \qquad i=\psi, \xi \ .
\end{eqnarray}
The  ultraviolet and infrared cutoffs are introduced to tame the integral in the respective regions. A simple physical interpretation of these cutoffs is the following. At very high energies, at scales above $\Lambda_{U}$, the underlying theory flows to the ultraviolet fixed point and we have to abandon the description in terms of the composite operator. This immediately suggests that $\Lambda_{UV}  $ is naturally identified with $\Lambda_U$. The presence of the mass terms induce mass gaps, which are the quantities we are trying to determine. It is hence natural to identify t the infrared cutoff  $\Lambda_{IR} $ with  with a linear combination of $| \langle \mathcal{O}_{\psi} \rangle |^{\frac{1}{3-\gamma_{\psi}}}$ and $| \langle \mathcal{O}_{\xi} \rangle |^{\frac{1}{3-\gamma_{\xi}}}$. We take simply $ \Lambda_{IR} \sim | \langle \mathcal{O}_{\psi} \rangle |^{\frac{1}{3-\gamma_{\psi}}} + | \langle \mathcal{O}_{\xi} \rangle |^{\frac{1}{3-\gamma_{\xi}}}$. In this way if any of the two condensates vanishes the other will cut off the infrared divergence when needed.

 Recalling the relations $\langle \tilde{\psi}\psi \rangle \sim \Lambda_U^{\gamma_{\psi}} \langle \mathcal{O}_{\psi} \rangle$ and $\langle \tilde{\xi}\xi \rangle \sim \Lambda_U^{\gamma_{\xi}} \langle \mathcal{O}_{\xi} \rangle$ we find
the following dependences on the masses and anomalous dimensions of the two wanted condensates:

\emph{$0< \gamma_{\psi} <1$}:
\begin{eqnarray}
\langle \tilde{\psi} \psi \rangle &\propto& -m_{\psi}  \Lambda_U^2 \ ,\\
\langle \tilde{\xi} \xi \rangle & \propto & -m_{\xi} \Lambda_U^2 \ , \qquad \ \ \ \  \qquad  \qquad 0<\gamma_{\xi}<1 \ , \\
\langle \tilde{\xi} \xi \rangle & \propto & - m_{\xi} \Lambda_U^2 \log\left( \frac{\Lambda_U}{\Lambda_{IR}} \right)^2 \ , \qquad \gamma_{\xi} \rightarrow 1 \ , \\
\langle \tilde{\xi} \xi \rangle & \propto & - m_{\xi} \Lambda_{IR}^2 \left( \frac{\Lambda_U}{\Lambda_{IR}}\right)^{2\gamma_{\xi}} \ , \qquad 1<\gamma_{\xi}<2 \ .
\end{eqnarray}
where $\Lambda_{IR} $ is function of the condensates itself, i.e. $\Lambda_{IR} \left[ \langle \tilde{\xi} \xi \rangle,  \langle \tilde{\psi}  \psi \rangle \right] $ and one has to solve the last two equations numerically.

\vskip .5cm

\emph{$ \gamma_{\psi} \rightarrow 1$}:
\begin{eqnarray}
\langle \tilde{\psi} \psi \rangle &\propto& -m_{\psi} \Lambda_U^2 \log \frac{\Lambda_U^2}{\Lambda_{IR}^2} \ , \\
\langle \tilde{\xi} \xi \rangle & \propto & -m_{\xi}\Lambda_U^2 \ , \qquad \qquad \qquad \qquad 0< \gamma_{\xi} <1 \ , \\
\langle \tilde{\xi} \xi \rangle & \propto & \frac{m_{\xi}}{m_{\psi}} \langle \tilde{\psi} \psi \rangle \ , \qquad \qquad \qquad \qquad \gamma_{\xi} \rightarrow 1  \ , \\
\langle \tilde{\xi} \xi \rangle & \propto & - m_{\xi} \Lambda_{IR}^2 \left( \frac{\Lambda_U}{\Lambda_{IR}}\right)^{2\gamma_{\xi}} \ , \qquad \qquad 1<\gamma_{\xi}<2 \ .
\end{eqnarray}

The case {$1< \gamma_{\psi} <2$} can be deduced from the case $0< \gamma_{\psi} <1$ with $\xi$ replaced by $\psi$ in all the equations.


The effect of the mass terms and the associated condensates is to break the conformal symmetry and some of the global symmetries. However there is a $U(1)$ global classical symmetry which is {\it always} broken by quantum corrections and we are interested on its effects on the conformal region. The original investigation for a single representation was first performed in \cite{Sannino:2008ha}.  In general the anomaly free global symmetry of the massless theory depends on the representation to which the two species of fermions belong and can be written as $G_{\psi} \times G_{\xi} \times U(1)$.
Only $\tilde{\psi}$ and $\psi$ are charged under $G_{\psi}$ while only $\tilde{\xi}$ and $\xi$ are charged under $G_{\xi}$. More specifically if $\tilde{\psi}$ and $\psi$ belong to a complex representation we have $G_{\psi}=SU(N_f(r_{\psi}))\times SU(N_f(r_{\psi}))\times U(1)$ while if they belong to a real or pseudoreal representation $G_{\psi}=SU(2N_f(r_{\psi}))$. The same is true for the other set of fermions $\tilde{\xi}$ and $\xi$.

What is relevant here it the abelian $U(1)$ symmetry under which all the fermions are charged. It is anomaly free provided the charge $Q$ of the fermions is
\begin{eqnarray}
Q\left[\tilde{\psi}, \psi  \right] = \frac{T(r_{\xi})}{N_f(r_{\psi})} \ , \qquad Q\left[ \tilde{\xi}, \xi \right] = - \frac{T(r_{\psi})}{N_f(r_{\xi})} \ .
\end{eqnarray}
Following \cite{Sannino:2008ha} we take the effects of instantons into account by adding the following term to the potential
\begin{eqnarray}\label{instantons}
&& \alpha N_f(r_{\psi}) \Lambda_U^4 \frac{\det \mathcal{O}_{\tilde{\psi}\psi} \left(\det \mathcal{O}_{\tilde{\xi}\xi} \right)^{\frac{T(r_{\xi})}{T(r_{\psi})}} }{\Lambda_U^{d_{\tilde{\psi}\psi} N_f(r_{\psi})} \Lambda_U^{d_{\tilde{\xi}\xi} \tilde{N}_f(r_{\xi})}} +
\beta N_f(r_{\xi}) \Lambda_U^4 \frac{\left(\det \mathcal{O}_{\tilde{\psi}\psi} \right)^{\frac{T(r_{\psi})  }{T(r_{\xi})} } \det \mathcal{O}_{\tilde{\xi}\xi} }{\Lambda_U^{d_{\tilde{\psi}\psi} \tilde{N}_f(r_{\psi})  } \Lambda_U^{d_{\tilde{\xi}\xi} N_f(r_{\xi})}} + \text{h.c.} \ ,
\end{eqnarray}
where we have defined $\tilde{N}_f(r_{\psi}) = N_f(r_{\psi}) \frac{T(r_{\psi})}{T(r_{\xi})}$ and $\tilde{N}_f(r_{\xi}) = N_f(r_{\xi}) \frac{T(r_{\xi})}{T(r_{\psi})}$. The new term in the Lagrangian is invariant under the intact global symmetry and in the limit $N_f(r_{\psi}) \rightarrow 0 $ (or $N_f(r_{\xi}) \rightarrow 0 $) we recover the instanton induced effective Lagrangian operator introduced for the single representation case in \cite{Sannino:2008ha}. Adding the above instanton induced term to \eqref{potential} and making the standard ansatz $\langle \mathcal{O}_{\tilde{\psi}\psi} \rangle = \langle \mathcal{O}_{\psi} \rangle \bf{1}_{N_f(r_{\psi}) \times N_f(r_{\psi})} $ and $\langle \mathcal{O}_{\tilde{\xi}\xi} \rangle = \langle \mathcal{O}_{\xi} \rangle \bf{1}_{N_f(r_{\xi}) \times N_f(r_{\xi})} $ the extrema of the potential are at:
\begin{eqnarray}
\langle \mathcal{O}_{\psi} \rangle &=&  - \mathcal{F}_{d_{\tilde{\psi}\psi}} \left[ \bar{N}_{\psi} + \bar{A}_{\psi} \langle \overline{\mathcal{O}}_{\psi} \rangle^{N_f(r_{\psi})-1}\langle \overline{\mathcal{O}}_{\xi} \rangle^{\tilde{N}_f(r_{\xi})} + \bar{B}_{\psi}  \langle \overline{\mathcal{O}}_{\psi} \rangle^{\tilde{N}_f(r_{\psi})-1} \langle \overline{\mathcal{O}}_{\xi} \rangle^{N_f(r_{\xi})}  \right]  \Omega_{\psi} \ ,  \\
\langle \mathcal{O}_{\xi} \rangle &=&  - \mathcal{F}_{d_{\tilde{\xi}\xi}} \left[  \bar{N}_{\xi} + \bar{A}_{\xi}   \langle \overline{\mathcal{O}}_{\psi} \rangle^{N_f(r_{\psi})}  \langle \overline{\mathcal{O}}_{\xi} \rangle^{\tilde{N}_f(r_{\xi})-1}     + \bar{B}_{\xi}  \langle \overline{\mathcal{O}}_{\psi} \rangle^{\tilde{N}_f(r_{\psi})} \langle \overline{\mathcal{O}}_{\xi} \rangle^{N_f(r_{\xi})-1}  \right] \Omega_{\xi} \ ,
\end{eqnarray}
where we have taken the limit $\Delta_{\psi}^2,\Delta_{\xi}^2 \rightarrow 0$. Also the coefficients are
\begin{eqnarray}
\bar{N}_i &=& \bar{m}_i \Lambda_U^{\gamma_i} N_f(r_i) \ ,\\
\bar{A}_i &=&  \bar{\alpha}  N_f(r_{\psi}) N_f(r_{i}) \frac{T(r_{i})}{T(r_{\psi})}   \Lambda_U^{4-d_{\tilde{\psi}\psi} N_f(r_{\psi}) - d_{\tilde{\xi}\xi} \tilde{N}_f(r_{\xi})} \ , \\
\bar{B}_i &=& \bar{\beta} N_f(r_{i}) N_f(r_{\xi}) \frac{T(r_{i})}{T(r_{\xi})}   \Lambda_U^{4-d_{\tilde{\xi}\xi} N_f(r_{\xi}) - d_{\tilde{\psi}\psi} \tilde{N}_f(r_{\psi})  } \ ,
\end{eqnarray}
where $i=\psi,\xi$. We shall solve the above two coupled equations in various limits. We, of course, recover the previous results when there is no instanton contribution $\bar{A}_i = \bar{B}_i =0,\ i=\psi,\xi $
\begin{eqnarray}
\langle \mathcal{O}_{\psi} \rangle = -\mathcal{F}_{d_{\tilde{\psi}\psi}} \bar{N}_{\psi} \Omega_{\psi} \left[ \Lambda_{UV},\Lambda_{IR} \right] \ , \qquad \langle \mathcal{O}_{\xi} \rangle = -\mathcal{F}_{d_{\tilde{\xi}\xi}} \bar{N}_{\xi} \Omega_{\xi} \left[ \Lambda_{UV},\Lambda_{IR} \right] \ .
\end{eqnarray}
Another interesting limit is when the instanton terms dominate and the extrema equations simplify to:
\begin{eqnarray}
0 &=&  \bar{N}_{\psi} + \bar{A}_{\psi} \langle \overline{\mathcal{O}}_{\psi} \rangle^{N_f(r_{\psi})-1}\langle \overline{\mathcal{O}}_{\xi} \rangle^{\tilde{N}_f(r_{\xi})} + \bar{B}_{\psi}  \langle \overline{\mathcal{O}}_{\psi} \rangle^{\tilde{N}_f(r_{\psi})-1} \langle \overline{\mathcal{O}}_{\xi} \rangle^{N_f(r_{\xi})}     \ ,  \\
0 &=&    \bar{N}_{\xi} + \bar{A}_{\xi}   \langle \overline{\mathcal{O}}_{\psi} \rangle^{N_f(r_{\psi})}  \langle \overline{\mathcal{O}}_{\xi} \rangle^{\tilde{N}_f(r_{\xi})-1}     + \bar{B}_{\xi}  \langle \overline{\mathcal{O}}_{\psi} \rangle^{\tilde{N}_f(r_{\psi})} \langle \overline{\mathcal{O}}_{\xi} \rangle^{N_f(r_{\xi})-1}    \ .
\end{eqnarray}
These equations cannot be solved in general. Note that the factor $\mathcal{F}\Omega$ drops out. Whether the $\alpha$ or $\beta$ term will dominate depends on the specific representations. Using the fact that $d_{\tilde{\psi}\psi},d_{\tilde{\xi}\xi} > 1$ at the infrared fixed point the dimension of the $\alpha$ term is lower than the dimension of the $\beta$ term if $T(r_{\psi}) > T(r_{\xi})$. Therefore in the instanton dominated (ID) limit \cite{Sannino:2008pz} and assuming $T(r_{\psi}) > T(r_{\xi})$ we find the following approximate solutions
\begin{eqnarray}
\langle \tilde{\psi}\psi \rangle_{\text{ID}} & \propto & \left[ m_{\psi}^{ 1 - \tilde{N}_f(r_{\xi}) } m_{\xi}^{ \tilde{N}_f(r_{\xi}) } \Lambda_U^{ 3 \left( N_f(r_{\psi}) + \tilde{N}_f(r_{\xi}) \right) -4 } \right]^{ \frac{1}{N_f(r_{\psi}) + \tilde{N}_f(r_{\xi}) -1 } } \ , \\
\langle \tilde{\xi}\xi \rangle_{\text{ID}} & \propto & \left[ m_{\psi}^{N_f(r_{\psi})} m_{\xi}^{1- N_f(r_{\psi})} \Lambda_U^{ 3 \left( N_f(r_{\psi}) + \tilde{N}_f(r_{\xi}) \right) -4 } \right]^{ \frac{1}{N_f(r_{\psi}) + \tilde{N}_f(r_{\xi}) -1 }} \ .
\end{eqnarray}
The result above nicely reproduces the one in \cite{Sannino:2008pz} in case of a single representation and illustrates that one cannot neglect the instantons when the mass operator becomes large. Conversely at very small masses one can neglect the instanton contribution when analyzing the effects on the conformal dynamics also in the case of multiple representations. 

The effects of these terms are, nevertheless, important on the chiral dynamics since they are needed to give masses to pseudogoldstone bosons associated to the anomalous $U(1)$ symmetry when the theory is used to describe (near) conformal technicolor.

In the discussion above we restricted ourselves to consider complex representations for the two fermion species. This choice, unfortunately, automatically excludes a wide class of theories from the analysis. To address this issue we generalize the potential and the extrema conditions for the case of one real and one pseudoreal representation. This is the case of the UMT model.   
To be specific we shall take $r_{\psi}$ to be pseudoreal while $r_{\xi}$ is taken to be real. Then $\mathcal{O}_{\tilde{\psi}\psi}$ and $\mathcal{O}_{\tilde{\xi}\xi}$ become $2N_f(r_{\psi}) \times 2N_f(r_{\psi})$ and $2N_f(r_{\xi}) \times 2N_f(r_{\xi})$ matrices respectively. The relevant mass terms are
\begin{eqnarray}
-m_{\psi} \Lambda_U^{\gamma_{\psi}} \text{Tr}\left[ \mathcal{O}_{\tilde{\psi}\psi} E_{\psi} \right] -m_{\xi} \Lambda_U^{\gamma_{\xi}} \text{Tr}\left[ \mathcal{O}_{\tilde{\xi}\xi} E_{\xi} \right] + \text{h.c.}
\end{eqnarray}
while the instanton induced terms are
\begin{eqnarray}
\alpha N_f(r_{\psi}) \Lambda_U^4 \frac{\text{Pf}\ \mathcal{O}_{\tilde{\psi}\psi} \left(\det \mathcal{O}_{\tilde{\xi}\xi} \right)^{\frac{T(r_{\xi})}{2T(r_{\psi})}} }{\Lambda_U^{d_{\tilde{\psi}\psi} N_f(r_{\psi})} \Lambda_U^{d_{\tilde{\xi}\xi} \tilde{N}_f(r_{\xi})}} +
\beta N_f(r_{\xi}) \Lambda_U^4 \frac{\left(\text{Pf}\  \mathcal{O}_{\tilde{\psi}\psi} \right)^{\frac{2T(r_{\psi})  }{T(r_{\xi})} } \det \mathcal{O}_{\tilde{\xi}\xi} }{\Lambda_U^{2d_{\tilde{\psi}\psi} \tilde{N}_f(r_{\psi})  } \Lambda_U^{2d_{\tilde{\xi}\xi} N_f(r_{\xi})}} + \text{h.c.}
\end{eqnarray}
Here $\text{Pf}$ is the Pfaffian and
\begin{eqnarray}
E_{\psi} =
\left( \begin{array}{cc}
0_{N_f(r_{\psi}) \times N_f(r_{\psi})} & 1_{N_f(r_{\psi}) \times N_f(r_{\psi})} \\
-1_{N_f(r_{\psi}) \times N_f(r_{\psi})} & 0_{N_f(r_{\psi}) \times N_f(r_{\psi})}
\end{array} \right) \ , \qquad
E_{\xi} =
\left( \begin{array}{cc}
0_{N_f(r_{\xi}) \times N_f(r_{\xi})} & 1_{N_f(r_{\xi}) \times N_f(r_{\xi})} \\
1_{N_f(r_{\xi}) \times N_f(r_{\xi})} & 0_{N_f(r_{\xi}) \times N_f(r_{\xi})}
\end{array} \right)
\end{eqnarray}
Similar to the case of complex representations we define two flavor singlet operators via $\text{Tr}\left[ \mathcal{O}_{\tilde{\psi}\psi} E_{\psi} \right] = 2N_f(r_{\psi}) \mathcal{O}_{\psi}$ and $\text{Tr}\left[ \mathcal{O}_{\tilde{\xi}\xi} E_{\xi} \right] = 2N_f(r_{\xi}) \mathcal{O}_{\xi}$.  Extremizing the above potential and using $\langle \mathcal{O}_{\tilde{\psi}\psi} \rangle = - \langle \mathcal{O}_{\psi} \rangle E_{\psi}$ and $\langle \mathcal{O}_{\tilde{\xi}\xi} \rangle =  \langle \mathcal{O}_{\xi} \rangle E_{\xi}$ we find
\begin{eqnarray}
\langle \mathcal{O}_{\psi} \rangle &=&  - \mathcal{F}_{d_{\tilde{\psi}\psi}} \left[ \bar{N}_{\psi} + \bar{A}_{\psi} \langle \overline{\mathcal{O}}_{\psi} \rangle^{N_f(r_{\psi})-1}\langle \overline{\mathcal{O}}_{\xi} \rangle^{\tilde{N}_f(r_{\xi})} + \bar{B}_{\psi}  \langle \overline{\mathcal{O}}_{\psi} \rangle^{2\tilde{N}_f(r_{\psi})-1} \langle \overline{\mathcal{O}}_{\xi} \rangle^{2N_f(r_{\xi})}  \right]  \Omega_{\psi} \ ,  \\
\langle \mathcal{O}_{\xi} \rangle &=&  - \mathcal{F}_{d_{\tilde{\xi}\xi}} \left[  \bar{N}_{\xi} + \bar{A}_{\xi}   \langle \overline{\mathcal{O}}_{\psi} \rangle^{N_f(r_{\psi})}  \langle \overline{\mathcal{O}}_{\xi} \rangle^{\tilde{N}_f(r_{\xi})-1}     + \bar{B}_{\xi}  \langle \overline{\mathcal{O}}_{\psi} \rangle^{2\tilde{N}_f(r_{\psi})} \langle \overline{\mathcal{O}}_{\xi} \rangle^{2N_f(r_{\xi})-1}  \right] \Omega_{\xi} \ ,
\end{eqnarray}
with
\begin{eqnarray}
\bar{N}_i &=& 2\bar{m}_i \Lambda_U^{\gamma_i} N_f(r_i) \ ,\\
\bar{A}_i &=&  \bar{\alpha} \left( -1 \right)^{\frac{N_f(r_{\psi}) (N_f(r_{\psi}) - 1) + \tilde{N}_f(r_{\xi}) }{2}} N_f(r_{\psi}) N_f(r_{i}) \frac{T(r_{i})}{T(r_{\psi})}   \Lambda_U^{4-d_{\tilde{\psi}\psi} N_f(r_{\psi}) - d_{\tilde{\xi}\xi} \tilde{N}_f(r_{\xi})} \ , \\
\bar{B}_i &=& \bar{\beta} 2\left(-1 \right)^{\tilde{N_f(r_{\psi})} (N_f(r_{\psi})-1) + N_f(r_{\xi}) } N_f(r_{i}) N_f(r_{\xi}) \frac{T(r_{i})}{T(r_{\xi})}   \Lambda_U^{4-2d_{\tilde{\xi}\xi} N_f(r_{\xi}) - 2d_{\tilde{\psi}\psi} \tilde{N}_f(r_{\psi})  } \ ,
\end{eqnarray}
where $i=\psi,\xi$. Conclusions similar to the ones for complex representations can be drawn. At this point it is straightforward to generalize the equations above to the case of one complex and one (pseudo)real representation and investigate the effects of introducing a source of strong CP violation \cite{Sannino:2008pz} i.e. the $\theta$ angle. 

\section{Conclusions}
We have investigated the gauge dynamics of nonsupersymmetric SU(N) gauge theories featuring the simultaneous presence of fermionic matter transforming according to two distinct representations of the underlying gauge group. We constructed the conformal {\it house} and shown that the sides of the houses reduce to the previously investigated conformal windows. In other words the old conformal windows are now only part of the house. The actual conformal house can be smaller but cannot be bigger than the one we predicted here.

Our results can be used for walking technicolor as well as unparticle model building. An explicit example is the Ultra Minimal Walking technicolor model which makes use of two different matter representations and it has phenomenological applications relevant to both LHC physics and cosmology. 
%

\end{document}